# Analytical Solutions for Bending Of Fireworks and Similarities with the Solution of Electromagnetic Wave Diffraction

Fathan Akbar[(1)] and Mikrajuddin Abdullah[(2,a)]

(1)Faculty of Mathematics and Natural Sciences

Bandung Institute of Technology,

(2) Department of Physics, Bandung Institute of Technology

Bandung Institute of Technology

Jl. Ganesa 10 Bandung  40132, Indonesia

[(a)]Email: mikrajuddin@gmail.com

## ABSTRACT

In this paper we examine more deeply about the bending mechanism of rod-shaped fireworks which burned from the free end. We derived new analytic equations. Surprisingly, we obtained the bending patterns are similar to the cornu spiral. With a few simple steps we proved that positions of points throughout the fireworks are given by Fresnel integrals, $C(x)$ and $S(x)$, which are generally found in phenomena of electromagnetic wave diffraction. Although we deeply discussed bending of fireworks rods, however the proposed method is likely to explain any phenomena in nature related to an evolving length scale associated with some material that becomes progressively stiff or dry, such as the growth of resin exuded from trees.

**Keywords**: fireworks bending, Fresnel integrals, cornu spiral

# I. INTRODUCTION

Entering the *Eid* celebration day, many Muslim communities, especially children, lit fireworks at the night before the *Eid* day. This night is known as the *Takbeer* night. There are many types of fireworks available. One famous type is a rod-shaped firework as shown in Figure 1. The fireworks are made of a metal rod (stainless steel, aluminium alloy, or other metals) coated with a flammable material. The fireworks is held at one end and burnt at the free end, usually using a matches. The firework undergoes a process of combustion from the free end to the handle end up of flammable materials burned out.

There was an interesting phenomenon we have observed. If fireworks are positioned at different angles to the horizontal, the fireworks rod will bend at different patterns. Previously we have explained the bending profile numerically [1]. The simulation results can explain the measurement al results. Indeed, there are so many natural phenomena around us or everyday activities that can become interesting topics of physics researches [1-16]. For most people, such phenomena or activities could be just common things. But, for physicists such phenomena may contain a number of wonderful rules to be explained.

In this paper we extend the discussion and try to develop analytic solutions. Surprisingly, we found an unexpected phenomenon. We observed the bending profile of very long fireworks resembles cornu spirals. Thus, there are similarities between bending behavior of fireworks with diffraction phenomena, which has been well known to produce cornu spiral [17]. We proved that the coordinates of bent fireworks are given by fresnel integrals, which is well known as the solution diffraction phenomena.

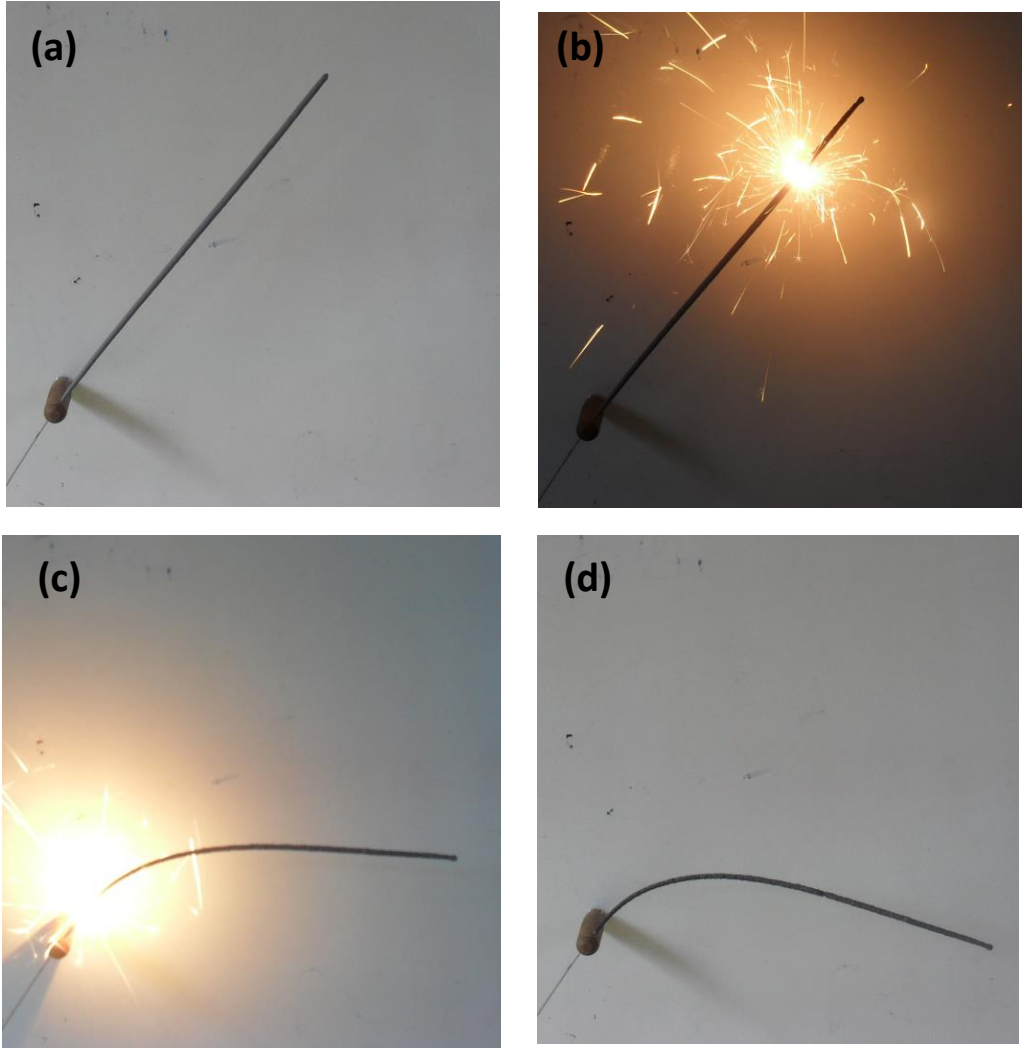

**Figure 1.** Profiles of fireworks during combustion process. (a) fireworks before being burned, (b) fireworks at initial burning process, (c) fireworks approaches the end of the combustion process, and (d) fireworks after complete combustion. We used fireworks made of aluminum alloy rod with a diameter of 1 mm, length of 40 cm, and positioned at a tilt angle of $60^o$.

## II. METHODS

Figure 2 (a) is the profile of the fireworks before being burned. The firework is rod shape, titled at $\theta_0$ to the horizontal. We divide the firework rod over $N$ small segments of the same length, $a = L/N$, with $L$ is the length of flammable material. Figure 2(b) is the profile of firework when the $k$-th segment is flaming, and the $(k+1)$-th segment just finished burning.

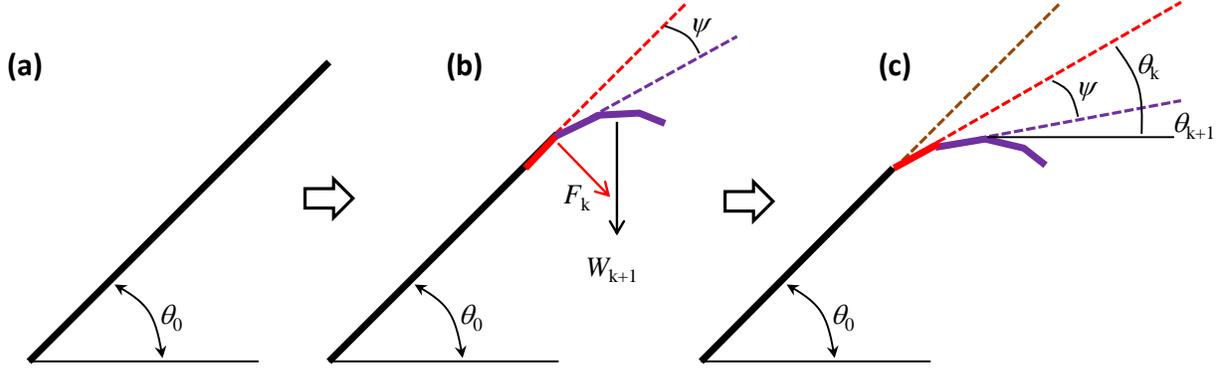

**Figure 2.** Illustration of the process of bending the firework rod. The firework rod is divided over $N$ segments of the same length. (a) firework before being burned. (b) firework after burnt up to $(k+1)$-th segment and (c) fireworks after burnt up to $k$-th segment.

The $k$-th segment supports the weight right part of the rod having a length $L-ka$. If $\lambda$ is the mass per unit length of the burnt part, then the weight of segments that have been burnt is

$$W_{k+1} = \lambda [L - ka] g \tag{1}$$

However, the force responsible for bending the $k$-th segment is the component of the force perpendicular to that segment, namely

$$F_k = W_{k+1} \cos \theta_0 \tag{2}$$

We hypothesize that the magnitude of deflection angle of the $k$-th segment is proportional to multiplication component of force perpendicular to the segment and segment length, and inversely proportional to the cross-section of segment. The same hypothesis we have used to develop a model to explain the mechanism of palm frond bending [18]. Mathematically, this hypothesis can be written

$$\psi = \alpha \frac{F_k a}{A} \qquad (3)$$

with $A$ is the cross-section of segment, and $\alpha$ is a proportional factor.

By observing Figure 2 (b) we obtain the difference in angles of the $k$-th and $(k+1)$-th segments

$$\theta_k - \theta_{k+1} = \psi \qquad (4)$$

Substitution of Eqs. (1) - (3) into the equation (4) we can then write

$$\theta_k - \theta_{k+1} = \alpha \frac{W_{k+1} a \cos\theta_0}{A} \qquad (5)$$

Furthermore, by replacement $\theta_k - \theta_{k+1} = -\Delta\theta$, $a = \Delta s$, $W_{k+1} = \lambda(L-s)g$, we can write Eq. (5) as $-\Delta\theta/\Delta s = \alpha\lambda(L-s)g\cos\theta_0/A$, or $d\theta/ds = -(\alpha\lambda g \cos\theta_0/A)(L-s)$, giving rise to the following solution

$$\theta(s) = \theta_0 - g\cos\theta_0 \int_0^s \frac{\alpha\lambda}{A}(L-s)ds \qquad (6)$$

Equation (6) applies to any fireworks rod, homogeneous or inhomogeneous, having a constant cross-section or cross-sections which are dependent on the distance from the fixed end. Especially for homogeneous fireworks, Eq. (6) becomes

$$\theta(s) = \theta_0 - \frac{\alpha\lambda g \cos\theta_0}{A}\left(Ls - \frac{1}{2}s^2\right) \qquad (7)$$

Next we will discuss some properties of the Eq. (7).

## III. DISCUSSION

From equation (7) we obtain the angle formed by the free end of the rod, ie at $s = L$, satisfies

$$\theta_f = \theta_0 - \frac{\alpha\lambda g \cos\theta_0}{2A}L^2 \qquad (8)$$

It appears that the angle of the free end changes with the square of the rod length and a linear function of the cosine of the fixed end angle.

Rod segment forms a horizontal angle if $\theta(s) = -n\pi$ is fulfilled, with $n = 0, 1, 2, \ldots$. The negative sign informs that the rod bending directs clockwise. By using equation (7), the distance from the fixed end of the segment forming a horizontal direction, $s_n$, satisfies

$$-n\pi = \theta_0 - \frac{\alpha\lambda g \cos\theta_0}{A}\left(Ls_n - \frac{1}{2}s_n^2\right)$$

or

$$\xi_n^2 - 2\xi_n + \frac{2A}{\alpha\lambda gL^2 \cos\theta_0}(n\pi + \theta_0) = 0$$

with $\xi_n = s_n/L$. The solution for $\xi_n$ is

$$\xi_n = 1 - \sqrt{1 - \frac{2A}{\alpha\lambda gL^2 \cos\theta_0}(n\pi + \theta_0)} \qquad (9)$$

From equation (9) we observe that the formation of horizontal directions is formed if $2A(n\pi + \theta_0)/\alpha\lambda gL^2 \cos\theta_0 \leq 1$ is fulfilled, or

$$n \leq \frac{1}{\pi}\left(\frac{\alpha\lambda gL^2 \cos\theta_0}{2A} - \theta_0\right) \qquad (10)$$

It appears that the location of horizontal direction increases when the rod is getting longer. We can determine the maximum number of locations horizontal direction, ie

$$n_{max} = round\left[\frac{1}{\pi}\left(\frac{\alpha\lambda gL^2 \cos\theta_0}{2A} - \theta_0\right)\right] \qquad (11)$$

where *round(x)* is an integer that equal to or smaller than *x*.

Conversely, the fireworks did not produce any horizontal direction when

$$\frac{1}{\pi}\left(\frac{\alpha\lambda gL^2 \cos\theta_0}{2A} - \theta_0\right) < 0$$

or

$$\frac{\cos\theta_0}{\theta_0} < \frac{2A}{\alpha\lambda gL^2} \tag{12}$$

Let us analyze the properties of inequality (12). If $L$ is very small, $2A/\alpha\lambda gL^2$ is very large so that $\cos\theta_0/\theta_0$ becomes very large. This condition is satisfied by $\theta_0 \to 0$. But, if $\theta_0 \to 0$ we can approximate $\cos\theta_0 \approx 1 - \theta_0^2/2$ such that inequality (12) can be approximated as

$$\frac{1 - \theta_0^2/2}{\theta_0} < \frac{2A}{\alpha\lambda gL^2} \tag{13}$$

or

$$\theta_0^2 + \frac{4A}{\alpha\lambda gL^2}\theta_0 - 2 > 0$$

giving raise to two possible solutions, ie

$$\theta_0 < -\frac{2A}{\alpha\lambda gL^2} - \sqrt{\left(\frac{2A}{\alpha\lambda gL^2}\right)^2 + 2} \tag{14a}$$

or

$$\theta_0 > -\frac{2A}{\alpha\lambda gL^2} + \sqrt{\left(\frac{2A}{\alpha\lambda gL^2}\right)^2 + 2} \qquad (14b)$$

Solution of (14a) gives to large negative value for $\theta_0$. However, since we have assumed that $\theta_0 \to 0$ this solution must be omitted. Thus the only solution that meets the requirement is (14b), which can be further approximated as

$$\theta_0 > \frac{\alpha\lambda gL^2}{2A} \qquad (15)$$

Conversely, if $L$ is very large, the right side of inequality (12) approaches zero. This condition is satisfied by $\theta_0 \to \pi/2$, and for this condition we can do the following approximation. $\cos\theta_0 = \cos[\pi/2 - (\pi/2 - \theta_0)] = \cos(\pi/2)\cos(\pi/2 - \theta_0) + \sin(\pi/2)\sin(\pi/2 - \theta_0) = \sin(\pi/2 - \theta_0)$. Since $\pi/2 - \theta_0 \to 0$ then $\sin(\pi/2 - \theta_0) \approx \pi/2 - \theta_0$. Therefore, in case of $\theta_0 \to \pi/2$ inequality (12) can be approximated as

$$\frac{\pi/2 - \theta_0}{\theta_0} < \frac{2A}{\alpha\lambda gL^2},$$

or

$$\theta_0 > \frac{\pi/2}{1 + \frac{2A}{\alpha\lambda gL^2}}$$

$$> \frac{\pi}{2}\left(1 - \frac{2A}{\alpha\lambda gL^2}\right) \qquad (16)$$

If the fireworks rod is very strong or has very large cross-section, the second term on the right-hand side of equation (7) approaches zero. It implies $\theta(s) \to \theta_0$ for all $s$. This means that the rod does not undergo any deflections or remain as straight line as before.

***Bending Profiles.*** Equation (7) expresses the bending angle as a function of distance from the fixed end. To figure out the bending profiles we need to determine the positions of all segments in $x$ and $y$ coordinates. We assume that the fixed end is located at $x = y = 0$. The position of the end of the first segment is $x_1 = x_0 + a\cos\theta(a)$, $y_1 = y_0 + a\sin\theta(a)$. The position of the end of the second segment is $x_2 = x_1 + a\cos\theta(2a)$, $y_2 = y_1 + a\sin\theta(2a)$. In general, the position of the end to the $k$-th segment is

$$x_k = x_{k-1} + a\cos\theta(ka), \tag{17a}$$

$$y_k = y_{k-1} + a\sin\theta(ka). \tag{17b}$$

Figure 3 is profiles of fireworks positioned at different tilt angles: (a) $80°$, (b) $60°$, and (c) $30°$. In the simulations we uses $L = 1$ m and $2\alpha\lambda g/A = 1.5$ m$^{-2}$. If the tilt angle increases (closer to vertical), the rod only deflect slightly. The deflection increases if the tilt angle approaches zero.

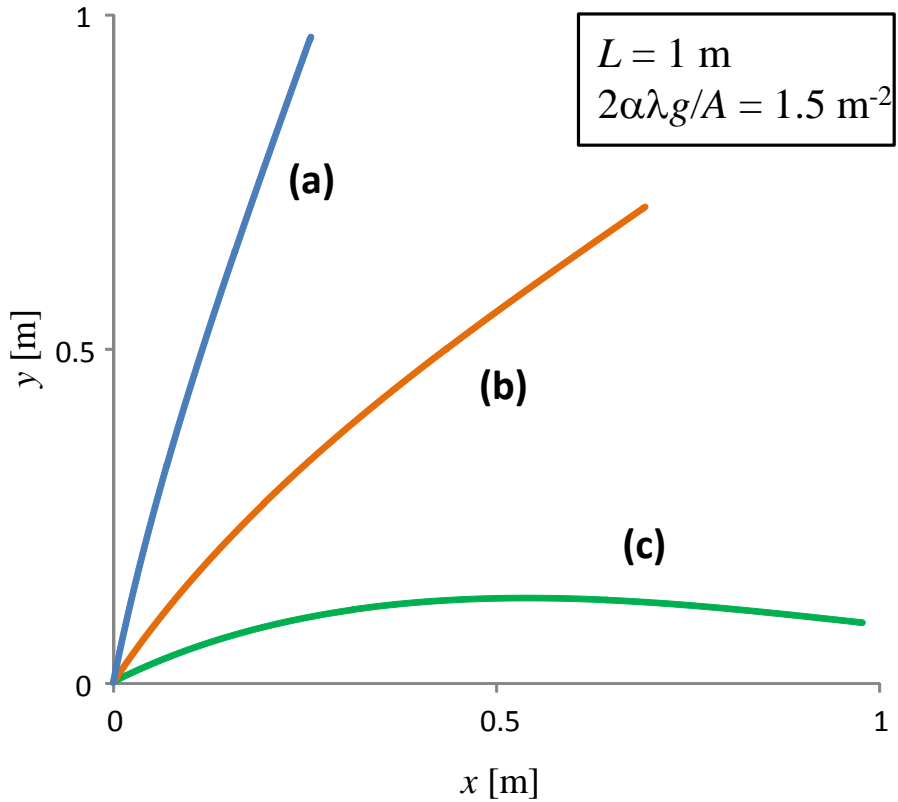

**Figure 3**. The result of calculation of the rod profiles that have been positioned at different tilt angles: (a) 80°, (b) 60°, and (c) 30°. In the calculations we uses $L = 1$ m and $2\alpha\lambda g/A = 1.5$ m$^{-2}$.

Figure 4 is profiles of rod bending at various lengths: (a) $L = 1$ m, (b) $L = 2$ m, (c) $L = 4$ m, and (d) $L = 6$ m. In all the simulations we used $2\alpha\lambda g/A = 1.5$ m$^{-2}$ and $\theta_0$ was fixed at 60°.

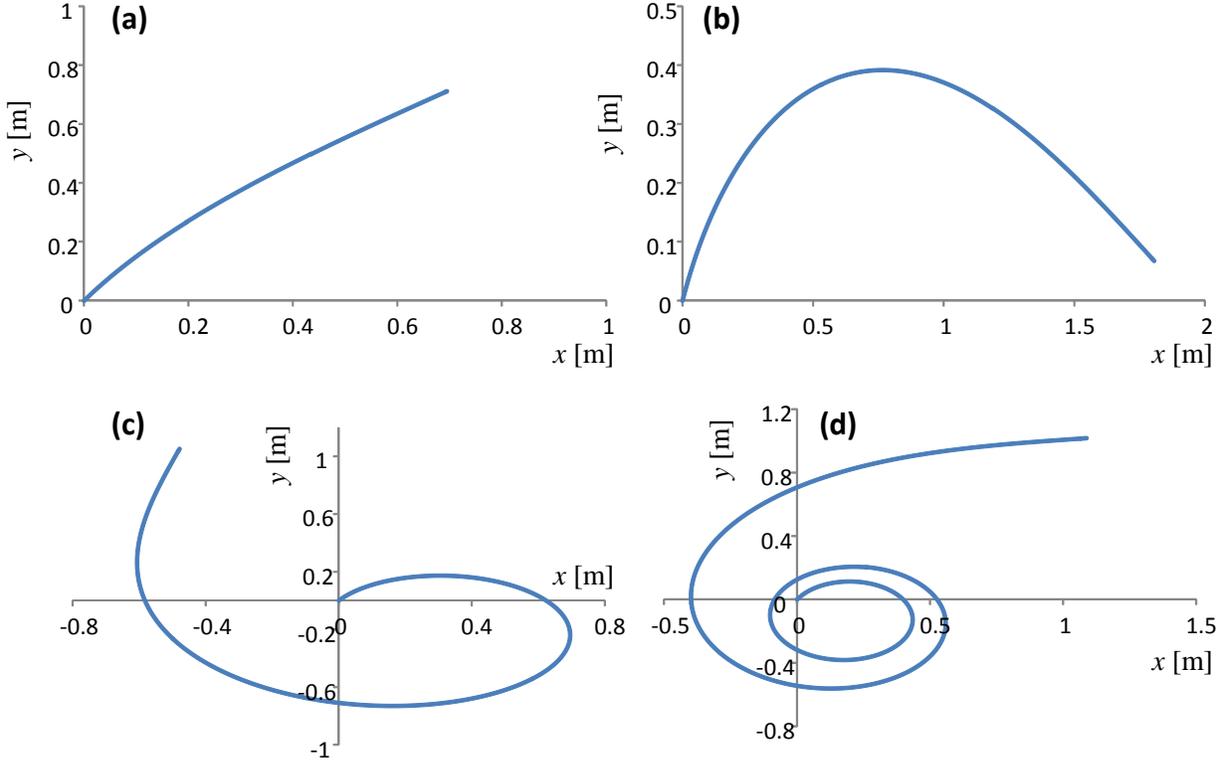

**Figure 4**. Profiles of rod bending at various lengths: (a) $L = 1$ m, (b) $L = 2$ m, (c) $L = 4$ m, and (d) $L = 6$ m. In all the simulations we use parameter $2\alpha\lambda g/A = 1.5$ m$^{-2}$ and the tilt angle is fixed at $60°$.

It appears in Figure 4 that if the rod length increases, the bending profile is more likely a spiral. The bending closely resembles a cornu spiral, which are widely used in diffraction problems [17]. It is therefore interesting to examine further such bending behavior.

We can further process the equation (17a) and (17b) as follows. We write

$$\frac{x_k - x_{k-1}}{a} = \cos\theta(ka), \tag{17a}$$

$$\frac{y_k - y_{k-1}}{a} = \sin\theta(ka). \tag{17b}$$

Let us define $x_k - x_{k-1} = \Delta x$, $y_k - y_{k-1} = \Delta y$, $a = \Delta s$, $ka = s$, and taking $\Delta s \to 0$ so that we obtain the following equation pair

$$\frac{dx}{ds} = \cos\theta(s), \tag{18a}$$

$$\frac{dy}{ds} = \sin\theta(s). \tag{18b}$$

Substituting $\theta(s)$ from Eq. (7) into Eqs. (18a) and (18b) resulting

$$\frac{dx}{ds} = \cos\left(\theta_0 - \frac{\alpha\lambda g \cos\theta_0}{A}\left(Ls - s^2/2\right)\right)$$

$$= \cos\left(\theta_0 - \beta L^2 + \beta(s-L)^2\right) \tag{19a}$$

and

$$\frac{dy}{ds} = \sin\left(\theta_0 - \beta L^2 + \beta(s-L)^2\right) \tag{19b}$$

with

$$\beta = \frac{\alpha\lambda g \cos\theta_0}{2A} \tag{20}$$

By using trigonometry identity then we can write

$$\frac{dx}{ds} = \cos(\theta_0 - \beta L^2)\cos(\beta(s-L)^2) - \sin(\theta_0 - \beta L^2)\sin(\beta(s-L)^2) \tag{21a}$$

$$\frac{dy}{ds} = \sin(\theta_0 - \beta L^2)\cos(\beta(s-L)^2) + \cos(\theta_0 - \beta L^2)\sin(\beta(s-L)^2) \tag{21b}$$

Thus, we get the expressions for $x(s)$ and $y(s)$ as follows

$$x(s) = \cos(\theta_0 - \beta L^2)\int_0^s \cos(\beta(s'-L)^2)ds' - \sin(\theta_0 - \beta L^2)\int_0^s \sin(\beta(s'-L)^2)ds' \tag{22a}$$

$$y(s) = \sin(\theta_0 - \beta L^2)\int_0^s \cos(\beta(s'-L)^2)ds' + \cos(\theta_0 - \beta L^2)\int_0^s \sin(\beta(s'-L)^2)ds' \tag{22b}$$

where we have selected $x(0) = y(0) = 0$.

To complete the integrals (22a) and (22b) we transforms $\beta(s'-L)^2 = t^2$ so that $ds' = \beta^{-1/2}dt$. Thus we can write

$$\int \cos(\beta(s'-L)^2)ds' = \beta^{-1/2}\int \cos(t^2)dt$$

$$= \beta^{-1/2}C(t) + A_1$$

$$= \beta^{-1/2}C(\sqrt{\beta}(s'-L)) + A_1$$

$$\int_0^s \cos(\beta(s'-L)^2)ds' = \beta^{-1/2}\{C(\sqrt{\beta}(s-L)) - C(\sqrt{\beta}(-L))\} \tag{23a}$$

In the same process we get

$$\int_0^s \sin(\beta(s'-L)^2)ds' = \beta^{-1/2}\{S(\sqrt{\beta}(s-L))-S(\sqrt{\beta}(-L))\} \qquad (23b)$$

where $C(x) = \int_0^x \cos(t^2)dt$ and $S(x) = \int_0^x \sin(t^2)dt$ are Fresnel integrals.

Both $C(x)$ and $S(x)$ are an odd functions so $C(-x) = -C(x)$ and $S(-x) = -S(x)$ and we can write

$$\int_0^s \cos(\beta(s'-L)^2)ds' = \beta^{-1/2}\{C(\sqrt{\beta}L)-C(\sqrt{\beta}(L-s))\} \qquad (24a)$$

$$\int_0^s \sin(\beta(s'-L)^2)ds' = \beta^{-1/2}\{S(\sqrt{\beta}L)-S(\sqrt{\beta}(L-s))\} \qquad (24b)$$

Finally we get the following equation for rod profile

$$\beta^{1/2}x(s) = \cos(\theta_0 - \beta L^2)\{C(\sqrt{\beta}L)-C(\sqrt{\beta}(L-s))\}$$
$$-\sin(\theta_0 - \beta L^2)\{S(\sqrt{\beta}L)-S(\sqrt{\beta}(L-s))\} \qquad (25a)$$

$$\beta^{1/2}y(s) = \sin(\theta_0 - \beta L^2)\{C(\sqrt{\beta}L)-C(\sqrt{\beta}(L-s))\}$$
$$+\cos(\theta_0 - \beta L^2)\{S(\sqrt{\beta}L)-S(\sqrt{\beta}(L-s))\} \qquad (25b)$$

From Eqs. (25a) and (25b) we get the coordinates of the rod free end as

$$\beta^{1/2}x(L) = \cos(\theta_0 - \beta L^2)C(\sqrt{\beta}L)-\sin(\theta_0 - \beta L^2)S(\sqrt{\beta}L) \qquad (26a)$$

$$\beta^{1/2} y(L) = \sin(\theta_0 - \beta L^2) C(\sqrt{\beta} L) + \cos(\theta_0 - \beta L^2) S(\sqrt{\beta} L) \qquad (26b)$$

Let us look back at Eqs. (25a) and (25b). We inspect the condition when $\beta \to 0$. By considering Eq. (20), this condition can be achieved if $\theta_0 \to \pi/2$ or either $\lambda$ or $\alpha$ is very small, or $A$ is very large. By using the properties of fresnel integral, $C(x) \approx x - x^5/10 \approx x$ and $S(x) \approx x^3/3 \approx 0$ if $x \to 0$ then we can approximate Eqs. (25a) and (25b) as follows

$$\beta^{1/2} x(s) \approx \cos(\theta_0 - \beta L^2) \{\sqrt{\beta} L - \sqrt{\beta}(L-s)\} - \sin(\theta_0 - \beta L^2)\{0 - 0\}$$

$$\approx \cos(\theta_0 - \beta L^2)\{\sqrt{\beta} s\}$$

or

$$x(s) \approx s \cos(\theta_0 - \beta L^2) \qquad (27a)$$

In the same process we obtain

$$y(s) \approx s \sin(\theta_0 - \beta L^2) \qquad (27b)$$

As mentioned above, $\beta \to 0$ can also be achieved by positioning the fireworks nearly vertical or $\theta_0 \to \pi/2$. At this condition, $\theta_0 \gg \beta L^2$ and further approximations that can be performed to Eqs. (27a) and (27b) are

$$x(s) \approx s[\cos\theta_0 \cos\beta L^2 + \sin\theta_0 \sin\beta L^2]$$

$$\approx s\beta L^2 \qquad (28a)$$

$$y(s) \approx s[\sin\theta_0 \cos\beta L^2 - \cos\theta_0 \sin\beta L^2]$$

$$\approx s(1 - \beta^2 L^4/2) \qquad (28b)$$

Conversely, if $\sqrt{\beta L} \gg 1$, we can make the following approximation. We use the properties of fresnel integral at $x \to \infty$, ie

$$S(x) \approx \sqrt{\frac{\pi}{2}} \left\{ \frac{sign(x)}{2} - [1 + O(x^{-4})] \left( \frac{\cos(x^2)}{x\sqrt{2\pi}} + \frac{\sin(x^2)}{x^3\sqrt{8\pi}} \right) \right\} \quad (29a)$$

$$C(x) \approx \sqrt{\frac{\pi}{2}} \left\{ \frac{sign(x)}{2} - [1 + O(x^{-4})] \left( \frac{\sin(x^2)}{x\sqrt{2\pi}} - \frac{\cos(x^2)}{x^3\sqrt{8\pi}} \right) \right\} \quad (29b)$$

With $sign(x) = -1, 0, 1$ for $x<0$, $x = 0$, and $x>0$, respectively. Because in our case $x$ is always positive, $sign(x) = 1$ so that

$$S(x) \approx \sqrt{\frac{\pi}{2}} \left\{ \frac{1}{2} - [1 + O(x^{-4})] \left( \frac{\cos(x^2)}{x\sqrt{2\pi}} + \frac{\sin(x^2)}{x^3\sqrt{8\pi}} \right) \right\}$$

$$\approx \sqrt{\frac{\pi}{2}} \left\{ \frac{1}{2} - \frac{\cos(x^2)}{x\sqrt{2\pi}} \right\} \quad (30a)$$

$$C(x) \approx \sqrt{\frac{\pi}{2}} \left\{ \frac{1}{2} - \frac{\sin(x^2)}{x\sqrt{2\pi}} \right\} \quad (30b)$$

At locations around the fixed end where $s \ll L$ so that $L-s \gg 1$, we get the following approximation

$$C(\sqrt{\beta}L) - C(\sqrt{\beta}(L-s)) \approx \sqrt{\frac{\pi}{2}} \left\{ \frac{1}{2} - \frac{\sin(\beta L^2)}{\sqrt{\beta L}\sqrt{2\pi}} \right\} - \sqrt{\frac{\pi}{2}} \left\{ \frac{1}{2} - \frac{\sin(\beta(L-s)^2)}{\sqrt{\beta}(L-s)\sqrt{2\pi}} \right\}$$

$$\approx \frac{1}{2\sqrt{\beta}} \left\{ \frac{\sin(\beta(L-s)^2)}{(L-s)} - \frac{\sin(\beta L^2)}{L} \right\} \quad (31a)$$

$$S\left(\sqrt{\beta}L\right) - S\left(\sqrt{\beta}(L-s)\right) \approx \sqrt{\frac{\pi}{2}}\left\{\frac{1}{2} - \frac{\cos(\beta L^2)}{\sqrt{\beta L}\sqrt{2\pi}}\right\} - \sqrt{\frac{\pi}{2}}\left\{\frac{1}{2} - \frac{\cos(\beta(L-s)^2)}{\sqrt{\beta(L-s)}\sqrt{2\pi}}\right\}$$

$$\approx \frac{1}{2\sqrt{\beta}}\left\{\frac{\cos(\beta(L-s)^2)}{(L-s)} - \frac{\cos(\beta L^2)}{L}\right\} \qquad (31b)$$

By using Eqs. (31a) and (31b), Eqs. (25a) and (25b) become

$$\beta^{1/2}x(s) \approx \frac{\cos(-\beta L^2)}{2\sqrt{\beta}}\left\{\frac{\sin(\beta(L-s)^2)}{(L-s)} - \frac{\sin(\beta L^2)}{L}\right\}$$

$$-\frac{\sin(-\beta L^2)}{2\sqrt{\beta}}\left\{\frac{\cos(\beta(L-s)^2)}{(L-s)} - \frac{\cos(\beta L^2)}{L}\right\}$$

$$= \frac{\cos(\beta L^2)}{2\sqrt{\beta}}\left\{\frac{\sin(\beta(L-s)^2)}{(L-s)} - \frac{\sin(\beta L^2)}{L}\right\}$$

$$+ \frac{\sin(\beta L^2)}{2\sqrt{\beta}}\left\{\frac{\cos(\beta(L-s)^2)}{(L-s)} - \frac{\cos(\beta L^2)}{L}\right\}$$

or

$$2\beta x(s) = \left\{\frac{\cos(\beta L^2)\sin(\beta(L-s)^2)}{(L-s)} - \frac{\cos(\beta L^2)\sin(\beta L^2)}{L}\right\}$$

$$+ \left\{\frac{\sin(\beta L^2)\cos(\beta(L-s)^2)}{(L-s)} - \frac{\sin(\beta L^2)\cos(\beta L^2)}{L}\right\}$$

$$= \frac{\sin(\beta L^2 + \beta(L-s)^2)}{(L-s)} - \frac{\sin(2\beta L^2)}{L} \qquad (32)$$

Since $L \gg 1$, $L-s \gg 1$, $\left|\sin(2\beta L^2)\right| \leq 1$, and $\left|\sin(\beta L^2 + \beta(L-s)^2)\right| \leq 1$ we have $x(s) \approx 0$. With a similar analysis we also obtain $y(s) \approx 0$.

Now we inspect the location near the free end where L-s << 1 so that $\sqrt{\beta}(L-s) \ll 1$. In this condition we can approximate $C(\sqrt{\beta}(L-s)) \approx \sqrt{\beta}(L-s)$, $S(\sqrt{\beta}(L-s)) \approx 0$, $C(\sqrt{\beta}L) \approx (1/2)\sqrt{\pi/2}$, and $S(\sqrt{\beta}L) \approx (1/2)\sqrt{\pi/2}$. Thus, Eqs. (25a) and (25b) can be rewritten as

$$\beta^{1/2} x(s) \approx \cos(-\beta L^2) \left\{ \frac{1}{2}\sqrt{\frac{\pi}{2}} - \sqrt{\beta}(L-s) \right\}$$

or

$$x(s) \approx \frac{1}{2}\sqrt{\frac{\pi}{2\beta}} \cos(\beta L^2) - (L-s)\cos(\beta L^2) \qquad (33)$$

Using a similar process we get

$$y(s) \approx -\frac{1}{2}\sqrt{\frac{\pi}{2\beta}} \sin(\beta L^2) + (L-s)\sin(\beta L^2) \qquad (34)$$

The analytical solutions we have derived, although have been deeply discussed for bending of fireworks rods, however it is likely to explain any phenomena in nature related to an evolving length scale associated with some material that becomes progressively stiff or dry, such as the growth of resin exuded from trees.

# CONCLUSION

We have derived analytical solutions to explain bending of fireworks rod or other rods that were soften sequentially from the free end. For very long rods, the bending profiles replicate the cornu spiral that commonly discussed in diffraction of electromagnetic wave, described by Fresnel integrals. Although we have been deeply discussed bending of fireworks rods, however the proposed method is likely to explain any phenomena in nature related to an evolving length scale associated with some material that becomes progressively stiff or dry, such as the growth of resin exuded from trees

ACKNOWLEDGEMENT

This work was supported by a research grant (No. 310y/I1.C01/PL/2015) from the Ministry of Research and Higher Education, Republic of Indonesia, 2015-2017.